\documentclass[twocolumn,showpacs,preprintnumbers,amsmath,aps]{revtex4}
\usepackage{graphicx}
\usepackage{dcolumn}
\usepackage{bm}
\usepackage{multirow}
\usepackage{tabularx}
\usepackage{color}
\usepackage{stmaryrd}
\begin{document}
\newcommand{\kvec}{\mbox{{\scriptsize {\bf k}}}}
\def\eq#1{(\ref{#1})}
\def\fig#1{Fig.\hspace{1mm}\ref{#1}}
\def\tab#1{table\hspace{1mm}\ref{#1}}
\title{
---------------------------------------------------------------------------------------------------------------\\

CaLi$_2$ superconductor under the pressure of 100 GPa: the thermodynamic critical field and the specific heat}
\author{R. Szcz{\c{e}}{\'s}niak}
\author{A.P. Durajski}\email{adurajski@wip.pcz.pl}
\author{P.W. Pach}
\affiliation{Institute of Physics, Cz{\c{e}}stochowa University of Technology, Ave. Armii Krajowej 19, 42-200 Cz{\c{e}}stochowa, Poland}
\date{\today} 
\begin{abstract}
In the paper, the thermodynamic properties of the CaLi$_2$ superconductor have been studied. It has been assumed the high value of the pressure: $p=100$ GPa. The thermodynamic critical field ($H_{C}$) and the specific heats for the superconducting and normal state 
($C^{S}$ and $C^{N}$) have been obtained. The calculations have been made within the framework of the Eliashberg approach for the wide range of the Coulomb pseudopotential: $\mu^{\star}\in\left<0.1,0.3\right>$. It has been shown that the low-temperature critical field and the specific heat jump at $T_{C}$ strongly decrease if $\mu^{\star}$ increases. The value of the dimensionless ratio $R_{C}\equiv\left(C^{S}-C^{N}\right)/C^{N}$ at the critical temperature differs significantly from the value predicted by the BCS model. In particular, the parameter $R_{C}$ decreases from $1.83$ to $1.73$ with the increase of the Coulomb pseudopotential value. In the paper, the interpolation formulas for the functions $T_{C}\left(\mu^{\star}\right)$ and $R_{C}\left(\mu^{\star}\right)$ have been also given.
\end{abstract}
\maketitle

%
%

Calcium under the compression exhibits the anomalous behavior associated with the structural transitions. In particular, Ca transforms from the $fcc$ to $bcc$ structure at $19.5$ GPa \cite{Olijnyk}. The simple cubic structure exist in the pressure ($p$) range from $32$ GPa to $113$ GPa. Above exist the Ca-IV phase \cite{Yabuuchi}. Several other higher pressure phases like Ca-V (at $p=139$ GPa \cite{Yabuuchi}), Ca-VI (at $p=158$ GPa \cite{Nakamoto}) and Ca-VII (at $p=210$ GPa \cite{Sakata}) have been also confirmed into the experimental way.

At $p=44$ GPa (the sc phase), the superconductivity with the critical temperature (T$_C$) close to $2$ K has been observed \cite{Dunn}. The critical temperature in calcium increases with the pressure and reaches the maximum value equal to $25$ K at $161$ GPa \cite{Yabuuchi01} - the highest value of the critical temperature for simple elemental superconductors.

Calcium can create the compounds with many other elements. The theoretical studies have predicted that the hydrogen-rich compound like CaH$_6$ ($p=150$ GPa) can be the high temperature superconductor with the extremely high value of the critical temperature: $T_{C}\in\left<220-235\right>$ K \cite{Wang}. However, this result is still not confirmed by the experimental data. 

On the other hand, the superconducting state in CaLi$_2$ has been experimentally observed at the pressure above $11$ GPa. From $11$ GPa to $40$ GPa, the critical temperature increases to maximum value equal to $13$ K, then $T_{C}$ slowly decreases to $9$ K at $81$ GPa \cite{Matsuoka2008}. 

The theoretical works suggest that CaLi$_2$ compound above $81$ GPa can also possesses the interesting superconducting properties. In particular, the critical temperature assumes the high value \cite{Xie}. Therefore in the presented paper, we have study the thermodynamic properties of the superconducting state of CaLi$_2$ at the pressure of $100$ GPa. 

In the considered case, the electron-phonon coupling constant is large: $\lambda=2.73$ \cite{Xie}; so the mean-field BCS theory can not be used \cite{Alexandrov}. For this reason, the calculation have been conducted within the framework of the Eliashberg formalism \cite{Eliashberg}. In particular, the thermodynamic critical field and the specific heats for the normal and superconducting state have been determined.

%
%

The Eliashberg equations on the imaginary axis can be written in the following form:
\begin{equation}
\label{r1}
\Delta_{n}Z_{n}=\frac{\pi}{\beta} \sum_{m}
\frac{K\left(n,m\right)-\mu^{\star}\theta\left(\omega_{c}-|\omega_{m}|\right)}{\sqrt{\omega_m^2+\Delta_m^2}}\Delta_{m}, 
\end{equation}
and
\begin{equation}
\label{r2}
Z_n=1+\frac {\pi}{\beta\omega _n }\sum_{m}\frac{K\left(n,m\right)}{\sqrt{\omega_m^2+\Delta_m^2}}\omega_{m},
\end{equation}
where the symbols $\Delta_{n}\equiv\Delta\left(i\omega_{n}\right)$ and $Z_{n}\equiv Z\left(i\omega_{n}\right)$ denote the order parameter and the wave function renormalization factor, respectively. The $n$-th Matsubara frequency is given by the expression: 
$\omega_{n}\equiv \frac{\pi}{\beta}\left(2n-1\right)$, where $\beta\equiv 1/k_{B}T$, and $k_{B}$ is the Boltzmann constant.

%
\begin{figure*}[!t]
\includegraphics[width=180mm]{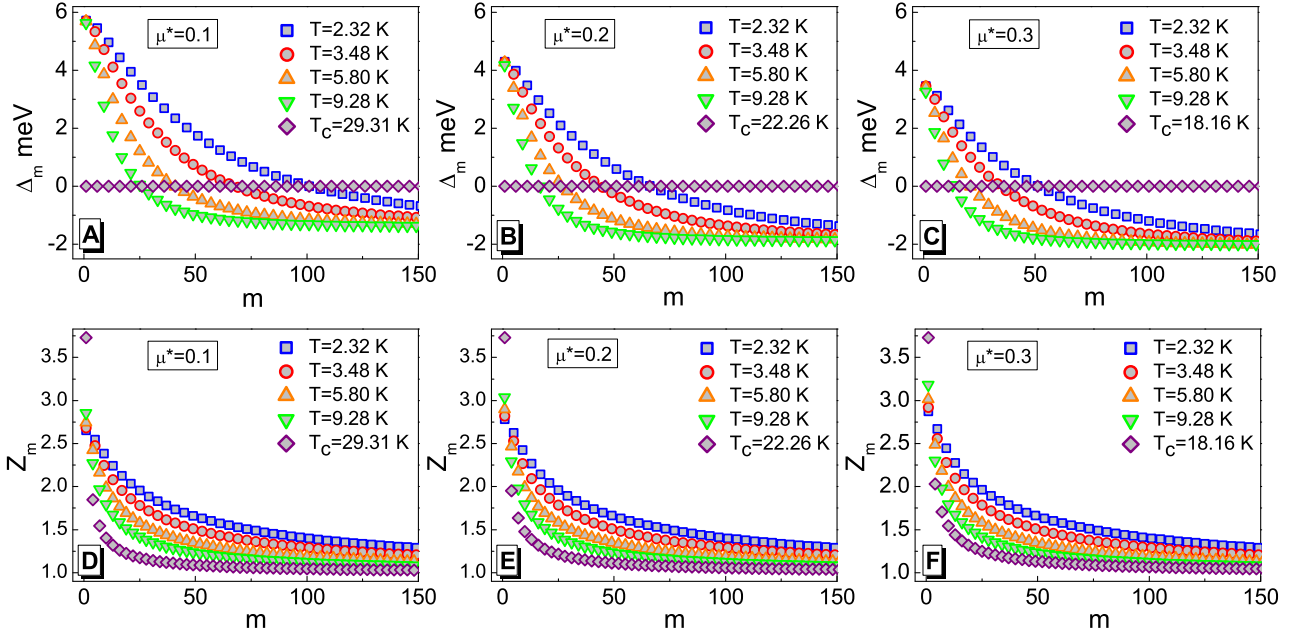}
\caption{The order parameter and wave function renormalization factor on the imaginary axis for the selected values of the temperature and the Coulomb pseudopotential. In the figures (A)-(C), the first $150$ values of the order parameter have been presented; (D)-(F) the $150$ values of the wave function renormalization factor.}
\label{f1}
\end{figure*}
%

The electron-phonon pairing kernel $K\left(n,m\right)$ is given by:
\begin{equation}
\label{r3}
K\left(n,m\right)\equiv 2\int_0^{\Omega_{\rm{max}}}d\Omega\frac{\Omega}
{\left(\omega_n-\omega_m\right)^2+\Omega ^2}\alpha^{2}F\left(\Omega\right),
\end{equation}
where the Eliashberg function ($\alpha^{2}F\left(\Omega\right)$) and the maximum phonon frequency ($\Omega_{\rm{max}}=113.1$ meV) has been determined 
in the paper \cite{Xie}. In particular, the crystal structure for ${\rm CaLi_{2}}$ compound has been obtained by using the {\it ab initio} evolutionary algorithm USPEX \cite{Oganov01}, \cite{Oganov02}, \cite{Glass}. This method has been also in detail discussed in the papers 
\cite{Ma01}, \cite{Oganov03}, \cite{Ma02}. The structural relaxations and the electronic properties calculations have been carried out using the density functional theory \cite{Hohenberg}, \cite{Kohn}, \cite{Perdew} (the VASP code \cite{Kresse}). The lattice dynamics and the electron-phonon coupling have been analyzed with help of the QUANTUM-ESPRESSO package \cite{Baroni}.

In the presented paper, the depairing electronic interaction has been described by the Coulomb pseudopotential ($\mu^{\star}$); the symbol $\theta$ denotes Heaviside unit function and $\omega_{c}$ is the cut-off frequency ($\omega_{c}=3\Omega_{\rm{max}}$). 

Due to the absence of the experimental value of the critical temperature, we assume the wide range of the Coulomb pseudopotential: $\mu^{\star}\in\left<0.1,0.3\right>$. This choice is connected with the fact that the value of the Coulomb pseudopotential in Ca and Li is high \cite{Radek01}, \cite{Luders}. For example: $\left[\mu^{\star}_{C}\right]^{\rm \left(Ca\right)}_{p=161 {\rm GPa}}=0.24$ and $\left[\mu^{\star}_{C}\right]^{\rm \left(Li\right)}_{p=22.3 {\rm GPa}}=0.22$ or $\left[\mu^{\star}_{C}\right]^{\rm \left(Li\right)}_{p=29.7 {\rm GPa}}=0.36$ \cite{Radek02}, \cite{Radek03}.

The Eliashberg equations have been solved for $2201$ Matsubara frequencies by using of the iteration method, described in the papers \cite{Radek04} and \cite{Radek05}. In the considered case, the functions $\Delta_{n}$ and $Z_{n}$ are stable for $T\geq T_{0}$, where the temperature $T_{0}$ is equal to $2.32$ K.

%
%

In \fig{f1}, the solutions of the Eliashberg equations on the imaginary axis have been presented. We have chosen the temperature range from T$_{0}$ to T$_C$ for the selected values of Coulomb pseudopotential. 

In \fig{f1} (A-C), the dependence of the order parameter's values on the successive Matsubara frequencies has been plotted. It is easy to see that the maximum value of the order parameter ($\Delta_{m=1}$) is decreasing with the growth of the temperature; the half-width of the $\Delta_{m}$ function becomes successively smaller. The last property suggests that together with the growth of the temperature, less of the Matsubara frequencies give relevant contribution to the Eliashberg equations.

In \fig{f1} (D-F), the wave function renormalization factor on the imaginary axis has been shown. In contrast to the maximum value of the order parameter, $Z_{m=1}$ slightly increases if the temperature increases. The half-width of the $Z_{m}$ function decreases with $T$. However, in comparison with the order parameter, this decrease is much slower.

We notice that the increase of the Coulomb pseudopotential strongly lowers the values of the order parameter and slightly influences on the wave function renormalization factor.

The thermodynamic properties of the superconducting phase depend on the free energy difference between the superconducting and the normal state ($\Delta F\equiv F^{S}-F^{N}$) \cite{Bardeen01}:

\begin{eqnarray}
\label{r4}
\frac{\Delta F}{\rho\left(0\right)}&=&-\frac{2\pi}{\beta}\sum_{n}
\left(\sqrt{\omega^{2}_{n}+\Delta^{2}_{n}}- \left|\omega_{n}\right|\right)\\ \nonumber
&\times&\left(Z^{S}_{n}-Z^{N}_{n}\frac{\left|\omega_{n}\right|}
{\sqrt{\omega^{2}_{n}+\Delta^{2}_{n}}}\right),
\end{eqnarray}
where $\rho\left(0\right)$ represents the value of the electron density of states at the Fermi surface. The symbols $Z^{S}_{n}$ and $Z^{N}_{n}$ denote the wave function renormalization factors for the superconducting and the normal state, respectively.

\begin{figure}[!h]
\includegraphics*[width=\columnwidth]{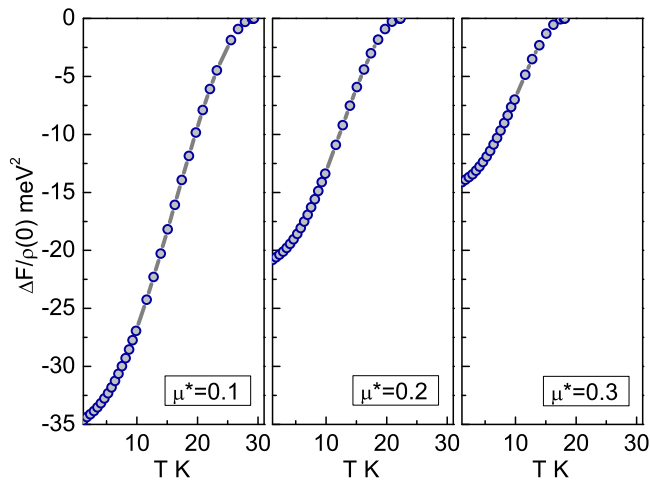}
\caption{The free energy difference between the superconducting and the normal state as a function of
the temperature for selected values of the Coulomb pseudopotential.}
\label{f2}
\end{figure}

In \fig{f2}, the dependence of $\Delta F/\rho\left(0\right)$ on the temperature for selected values of the Coulomb pseudopotential has been plotted. The negative values of the considered ratio prove that the superconducting state is thermodynamically stable to the critical temperature. Additionally, it can be noticed that $\Delta F\left(T_{0}\right)$ strongly decreases with $\mu^{\star}$:

\begin{equation}
\label{r5}
\left[\Delta F\left(T_{0}\right)\right]_{\mu^{\star}=0.3}/\left[\Delta F\left(T_{0}\right)\right]_{\mu^{\star}=0.1}\simeq 0.41.
\end{equation}
%
 
The thermodynamic critical field can be calculated in the following way \cite{Carbotte}:

\begin{equation}
\label{r6}
\frac{H_{C}}{\sqrt{\rho\left(0\right)}}=\sqrt{-8\pi\left[\Delta F/\rho\left(0\right)\right]}.
\end{equation}

\begin{figure}[!h]%
\includegraphics*[width=\columnwidth]{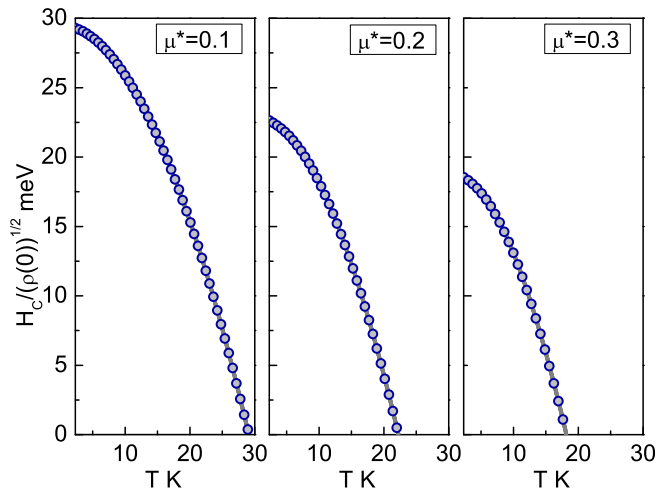}
\caption{The thermodynamic critical field as a function of the temperature for selected values
of the Coulomb pseudopotential.}
\label{f3}
\end{figure}

In \fig{f3}, the temperature dependence of the ratio $H_{C}/\sqrt{\rho\left(0\right)}$ has been shown. It is easy to see that the low temperature value of the critical field strongly decreases if the Coulomb pseudopotential increases:
\begin{equation}
\label{r7}
\left[H_{C}\left(T_{0}\right)\right]_{\mu^{\star}=0.3}/\left[H_{C}\left(T_{0}\right)\right]_{\mu^{\star}=0.1}\simeq 0.65.
\end{equation}

%
%

The specific heat difference between the superconducting and the normal state $\left(\Delta C\equiv C^{S}-C^{N}\right)$ should be calculated by using the expression:

\begin{equation}
\label{r8}
\frac{\Delta C\left(T\right)}{k_{B}\rho\left(0\right)}=-\frac{1}{\beta}\frac{d^{2}\left[\Delta F/\rho\left(0\right)\right]}{d\left(k_{B}T\right)^{2}}.
\end{equation}

Whereas, the specific heat in the normal state can be obtained from the formula: 
$\frac{C^{N}\left(T\right)}{ k_{B}\rho\left(0\right)}=\frac{\gamma}{\beta}$, where the Sommerfeld constant is given by: $\gamma\equiv\frac{2}{3}\pi^{2}\left(1+\lambda\right)$. 

%
\begin{figure}[!h]%
\includegraphics*[width=\columnwidth]{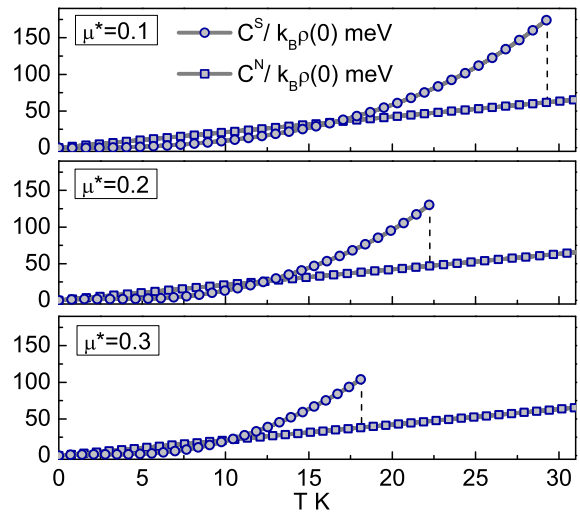}
\caption{The dependences of the specific heats in the superconducting and the normal states on the temperature for selected
values of the Coulomb pseudopotential. The vertical line indicates the position of specific heat's jump at T$_C$.}
\label{f4}
\end{figure}
%

In \fig{f4}, the form of the specific heats has been shown. 
Let us notice that at low temperatures, the specific heat in the superconducting state is exponentially suppressed. For the higher temperature, 
$C^ {S}$ rapidly increases and much exceed the value of $C^{N}$. At the critical temperature the characteristic jump has been observed.

Furthermore, we can see that at the critical temperature the specific heat's jump decreases with the growth of the Coulomb pseudopotential:

\begin{equation}
\label{r9}
\left[\Delta C\left(T_{C}\right)\right]_{\mu^{\star}=0.3}/\left[\Delta C\left(T_{C}\right)\right]_{\mu^{\star}=0.1}\simeq 0.59.
\end{equation}

The knowledge of the functions $C^{S}$ and $C^{N}$ enables the determination of the parameter:

\begin{equation}
\label{r10}
R_{C}\equiv\frac{\Delta C\left(T_{C}\right)}{C^{N}\left(T_{C}\right)}.
\end{equation}

For CaLi$_2$ compound under the pressure at $100$ GPa, we have obtained: $R_{C}\in\left<1.83,1.73\right>$, if $\mu^{\star}\in\left<0.1,0.3\right>$.
The above result strongly differs from the value predicted by the BCS model: $\left[R_{C}\right]_{{\rm BCS}}=1.43$ \cite{Bardeen02}.
The observed difference between the prediction of the BCS and Eliashberg theory is connected with the existence of the strong-coupling and retardation effects in the CaLi$_2$ compound. These effects are measured by the parameter: $k_{B}T_{C}/\omega_{\rm ln}$, where the symbol 
$\omega_{\rm ln}$ denotes the logarithmic phonon frequency \cite{AllenDynes}. In the BCS limit, we have: $k_{B}T_{C}/\omega_{\rm ln}\rightarrow 0$. On the other hand, in the framework of the Eliashberg theory, we can obtain: $k_{B}T_{C}/\omega_{\rm ln}\in\left<0.656,0.404\right>$ for $\mu^{\star}\in\left<0.1,0.3\right>$. We notice that the critical temperature has been calculated from the formula determined in the Appendix and $\omega_{{\rm ln}}=3.87$ meV.

In order to determine the value of $R_{C}$, it has been necessary to made the complicated calculations. For this reason, we present the convenient interpolation formula also for $R_{C}$ in order to reconstruct the precise numerical result: 

\begin{eqnarray}
\label{r11}
\frac{R_{C}}{\left[R_{C}\right]_{{\rm BCS}}}=
1+53\left(\frac{0.158k_{B}T_{C}}{\omega_{{\rm ln}}}\right)^{2}\ln\left(\frac{1.073\omega_{{\rm\ln}}}{k_{B}T_{C}}\right).
\end{eqnarray}
%

%
\begin{figure}[!h]%
\includegraphics*[width=\columnwidth]{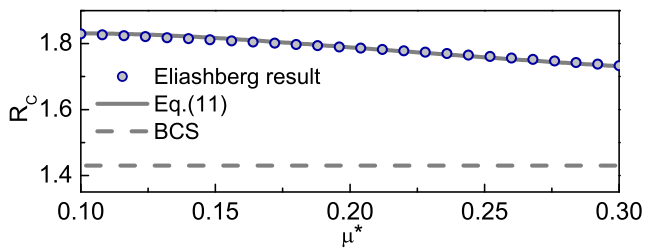}
\caption{The value of the ratio R$_{C}$ as a functions of the Coulomb pseudopotential. The circles represent the exact Eliashberg result. The solid line has been obtained on the basis of the expression \eq{r11}. The BCS result has been also shown.}
\label{f5}
\end{figure}
%

In Fig \ref{f5}, we have presented the ratio $R_{C}$ as a function of the Coulomb pseudopotential. It is easy to see that the formula \eq{r11} reproduces correctly the numerical data.

%
%

In summary, we have used the Eliashberg equations to calculate the thermodynamic critical field and the specific heats of CaLi$_2$ superconductor under the pressure at $100$ GPa. We have shown that the low-temperature value of the critical field and the value of the specific heat jump at $T_{C}$ strongly deceases if the Coulomb pseudopotential increases. The ratio $R_{C}$ for the characteristic values of the functions $C^{S}$ and $C^{N}$ can not be described correctly by the conventional BCS theory. In particular, $R_{C}\in\left<1.83,1.73\right>$, where $\mu^{\star}\in\left<0.1,0.3\right>$. The numerical results for the functions $T_{C}\left(\mu^{\star}\right)$ and $R_{C}\left(\mu^{\star}\right)$  have been supplemented by the interpolation formulas.

\vspace*{0.25cm}

{\bf Acknowledgments}

\vspace*{0.25cm}

The calculations have been based on the Eliashberg function sent to us by: Prof. Yanming Ma and Ph.D. Yu Xie to whom we are very thankful.\\
The authors are grateful to the Cz{\c{e}}stochowa University of Technology - MSK
CzestMAN for granting access to the computing infrastructure built in the
project No. POIG.02.03.00-00-028/08 "PLATON - Science Services Platform".
 

\vspace*{0.25cm}

{\bf Appendix}

\vspace*{0.25cm}

The interpolation formula for the critical temperature takes the form:

\renewcommand{\theequation}{A1} 
\begin{equation}
\label{A1}
k_{B}T_{C}=f_{1}f_{2}\frac{\omega_{{\rm ln}}}{1.45}\exp\left[\frac{-1.172\left(1+\lambda\right)}{\lambda-\mu^{\star}}\right],
\end{equation}
where $f_{1}$ and $f_{2}$ denote the strong-coupling correction function and the shape correction function, respectively \cite{AllenDynes}:

\begin{equation}
\label{A2}
f_{1}\equiv\left[1+\left(\frac{\lambda}{\Lambda_{1}}\right)^{\frac{3}{2}}\right]^{\frac{1}{3}} \qquad {\rm and,} \qquad
f_{2}\equiv 1+\frac{\left(\frac{\sqrt{\omega_{2}}}{\omega_{\rm{ln}}}-1\right)\lambda^{2}}{\lambda^{2}+\Lambda^{2}_{2}}.
\end{equation}
The quantities $\Lambda_{1}$ and $\Lambda_{2}$ have the form:
\begin{equation}
\label{A3}
\Lambda_{1}=2-2.6\mu^{\star}\qquad {\rm and,}\qquad
\Lambda_{2}=0.31+1.41\mu^{\star}\left(\frac{\sqrt{\omega_{2}}}{\omega_{\ln}}\right).
\end{equation}
The parameter $\sqrt{\omega_{2}}$ is equal to $34.95$ meV.

The formula \eq{A1} has been prepared basing on $300$ exact values of the function $T_{C}\left(\mu^{\star}\right)$. We have used the last square method and we have considered the range of $\mu^{\star}$ from $0.1$ to $0.3$.

%

%
\end{document}